\documentclass[doublecolumn,10pt,twoside]{IEEEtran}

\newcommand{\RNum}[1]{\lowercase\expandafter{\romannumeral #1\relax}}
\usepackage{amsthm,amsmath,amssymb,esint,color,graphicx,overpic,bbm}
\newcommand{\subparagraph}{}\usepackage[compact]{titlesec}
\usepackage{cite,hyperref}
\usepackage{array}
\usepackage{multirow}
\usepackage{tabu}
\usepackage{subcaption}
\usepackage{tabulary}
\theoremstyle{plain}

\theoremstyle{definition}
\theoremstyle{plain}

\theoremstyle{definition}

\providecommand{\definitionname}{Definition}

\providecommand{\lemmaname}{Lemma}
\providecommand{\theoremname}{Theorem}
\providecommand{\remarkname}{Remark}

\def\QED{\mbox{\rule[0pt]{1.3ex}{1.3ex}}}

\newcommand{\bm}[1]{{\boldsymbol {#1}}}

\usepackage{algorithmic}
\usepackage[ruled,vlined,linesnumbered]{algorithm2e}
\SetKwRepeat{Do}{do}{while} 

\begin{document}

\title{UAV Trajectory Optimization for Time Constrained Applications}

\newcounter{one}
\setcounter{one}{1}
\newcounter{two}
\setcounter{two}{2}
\author{
Emmanouil Fountoulakis,
Georgios S. Paschos, and Nikolaos Pappas
\vspace{0.1in}\\
\thanks{
E. Fountoulakis and N. Pappas are with the Department of Science and Technology, Link\"oping University, Norrk\"oping, Sweden. Emails: \{emmanouil.fountoulakis, nikolaos.pappas\}@liu.se.

G. S. Paschos is with Amazom.com, Luxembourg city, Luxembourg. Email: paschosg@amazon.com. The work of G. S. Paschos was done before joining Amazon.com.
	
The views and opinions expressed in this article are those of the authors and do not necessarily reflect the official policy or position of Amazon.com.} 
}

\maketitle

\addtolength{\floatsep}{-\baselineskip}
\addtolength{\dblfloatsep}{-\baselineskip}
\addtolength{\intextsep}{-\baselineskip}
\addtolength{\textfloatsep}{-\baselineskip}
\addtolength{\dbltextfloatsep}{-\baselineskip}
\addtolength{\abovedisplayskip}{0ex}
\addtolength{\belowdisplayskip}{0ex}
\addtolength{\abovedisplayshortskip}{0ex}
\addtolength{\belowdisplayshortskip}{0ex}
\setlength{\abovecaptionskip}{0ex}
\setlength{\belowcaptionskip}{0ex}

\begin{abstract}
In this paper, we consider a UAV flying over multiple locations and serves as many
users as possible within a given time duration. We study the
problem of optimal trajectory design, which we formulate as a
mixed-integer linear program. For large instances of the problem
where the options for trajectories become prohibitively many, we
establish a connection to the orienteering problem, and propose
a corresponding greedy algorithm. Simulation results show that
the proposed algorithm is fast and yields solutions close to the
optimal ones. The proposed algorithm can be used for trajectory
planning in content caching or tactical field operations. 

\end{abstract}

\section{Introduction}
The growing popularity of mobile devices and applications that require bandwidth hungry services has fueled an increase in mobile traffic  which in many cases, limits the ability of systems to offer high quality communications. 
For example, in peak demand hours, or during a popular event in  specific areas, communication systems become congested and service quality is critically impaired. In addition, applications that require operations in areas without infrastructure, called \textit{tactical field operations}, raise the need for fast and reliable reaction from the communication systems.
In such situations, it is desirable to have helpers to offload traffic from  congested networks or areas without infrastructure \cite{zhao2019caching}, \cite{baek2018design}.
Unmanned Aerial Vehicles (UAVs) can fly and serve congested network areas or specific areas that require urgently specific information. However, UAVs have limited flight time duration. Therefore, a UAV may not have the energy resources to visit all the areas. In this paper, our goal is to design an optimal trajectory in order to serve areas with higher emergency within a certain time.

 
Recently, research on UAVs that act as small base stations or caching helpers, has attracted a lot of interest \cite{zhao2019caching,baek2018design,zeng2016wireless}. The authors in \cite{cao2018mobile} consider a UAV that flies from one location to another and  accomplishes a certain amount of computation tasks. 
UAVs  are used as small cells with caches in \cite{lakiotakis2019joint}. Caching and UAV placement strategies are provided while considering limited UAV battery budget constraints. Authors in \cite{xu2018overcoming} consider that a UAV transmitting files to a group of ground terminals (GTs). Based on device-to-device (D2D) communications, GTs can share the files, received by the UAV, to their adjacent GTs when they are requested. In \cite{chen2017caching}, a proactive caching technique is considered. The authors propose a solution for UAVs deployment and caching content placement in order to maximize the quality-of-service (QoS). In \cite{samir2019trajectory}, the authors consider the   UAV deployment for data delivery in vehicular networks. \emph{To the best of our knowledge, there is no work that considers a UAV that flies over multiple areas with high importance and serve as many as possible within a certain time.}
The importance of each area is expressed with a score. We formulate an optimization problem whose solution provides a trajectory for the UAV for which the collected score is maximized. By drawing an analogy from the orienteering problem \cite{golden1987orienteering}, we prove that the problem is NP-hard. We provide a greedy algorithm that finds an approximate solution to the optimization problem in scalable manner.
\section{System model and Problem Formulation}
We consider a UAV that flies over multiple geographical areas and collects the corresponding scores. Location $i$ has a score which is denoted by $\lambda_{i}$ \footnote{E.g. the reward can play the role of user demand in that location.}. 
Our goal is to design a trajectory that maximizes the scores collected by the UAV.

\textbf{Trajectory selection.}
We consider that each location $i\in \mathcal{I}$ may be the barycenter (or centroid) of that area, or some central hotspot point. Let $x_i$ denote its coordinates on the plane. 
The UAV forms a \emph{trajectory} by visiting a subset of  locations in a specified order. 

\begin{figure}
	\centering
	\includegraphics[scale=0.335]{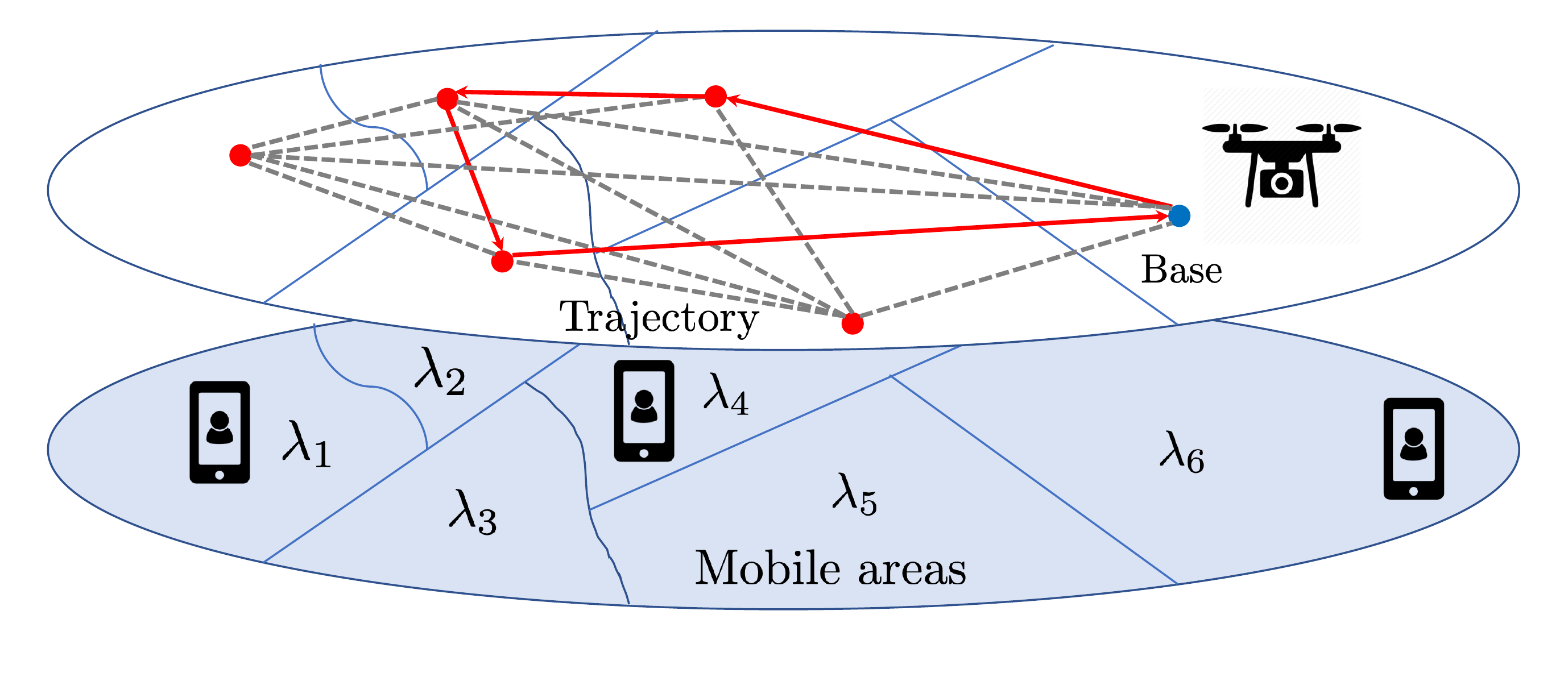}
	\caption{Illustration of our system model with six areas.}
	\label{fig:drone2}
\end{figure}

For two locations $i,j\in \mathcal{I}$, let $d_{ij}=d_{ji}\propto\|x_i-x_j\|^2$ be a \emph{distance}  that measures the amount of time it takes the UAV to move from one location to the other (for example the Euclidean distance of $x_i,x_j$ divided by the maximum velocity of the UAV).
Consider an undirectional complete graph $G=(\mathcal{I},E,\boldsymbol d)$, where  $E=\left\{\{i,j\}:i,j \in \mathcal{I}, i\neq j\right\}$ is the set of links connecting the locations, and for each link $\{i,j\}$ we have an associated distance $d_{ij}$ (or $d_{ji}$).
A {trajectory} $\mathcal{T}$ is a tour on graph $G$, i.e., an ordered set of nodes $\mathcal{T}\triangleq(i_1,i_2 \dots, i_k, i_1)$, such that the UAV visits the nodes in the described order and all nodes are visited once,  except $i_1$. An example of our system model is shown in Fig. \ref{fig:drone2}, where $\mathcal{T} = (6,4,2,3,6)$.

For each link $(i,j)\in E$, we introduce flow variables $f_{ij}\in\{0,1\}$, where $f_{ij}=1$ if and only if nodes $i,j$ appear in the trajectory $\mathcal{T}$. The flow that enters node $i$ must be equal with the one that goes out:
\begin{align}
 \sum_{j}f_{ji} =\sum_{j}f_{ij}=\left\{\begin{array}{ll}\label{eq: flow}
 1\text{,} & \text{if } i\in \mathcal{T} \\
 0\text{,} & \text{if } i\notin \mathcal{T} 
\end{array}
\right.,\quad \forall i\in \mathcal{I}.
\end{align}

 However, including only the constraints in (\ref{eq: flow}), we may produce tours that are not connected. In order to create solutions that do not contain disconnected tours, we introduce the following \emph{subtour elimination constraints}:
\begin{align}\label{cons: subtour}
	\sum_{(i,j) \in \mathcal{I}\text{, } i\neq j} f_{ij} \leq |\mathcal{S}| - 1 \text{, } \forall \mathcal{S} \subset \mathcal{I}\text{.}
\end{align}
For each nonempty subset $\mathcal{S}$ of the set of nodes $\mathcal{I}$, (\ref{cons: subtour}) ensures that the number of edges of $\mathcal{S}$ must be at most $|\mathcal{S}| - 1 $. Hence, (\ref{cons: subtour}) eliminates solutions with two or more disconnected subtours.
Note, that any non-zero integer flow $\boldsymbol f$ that satisfies \eqref{eq: flow} and \eqref{cons: subtour} is a tour, i.e., a path on graph $G$ that starts and ends at the same node. In case we need a trajectory that starts and ends at a specific node $s\in \mathcal{I}$, we can include $\sum_j f_{sj}=1$ as a constraint. 

The total time of the  trajectory $\mathcal{T}$ is denoted by $D$ and is equal to the sum of all traversed distances, $D=\sum_{(i,j)} d_{ij}f_{ij}$. 
A trajectory is called \emph{feasible} if its total time is no larger than a specified limit $D_{\max}$:
\begin{align}\label{cons: deadline}
	D = \sum_{(i,j)} d_{ij}f_{ij} \leq D_{\max}.
\end{align}

Our target is to compute a UAV trajectory that passes through the areas with highest scores in time less than $D_{\text{max}}$. To this end, we formulate the UAV Trajectory Design (UTD) problem as:

%

\begin{subequations}\label{optproblem: trajdesign}
	\begin{align}
		\max_{\boldsymbol{f}} & \sum_{(i,j)}f_{ij} \lambda_i \label{eq:obj1}\\
		\text{s.~t.}&\text{ }(\ref{eq: flow}), (\ref{cons: subtour}), (\ref{cons: deadline}), \\
	     & \text{ }\sum_{j}f_{ij} \leq 1\text{, } \forall i \in \mathcal{I} \label{eq: singleflowcon}\text{,} \\
	     &\text{ } \boldsymbol f\in \{0,1\}^{|\mathcal{I}|\times |\mathcal{I}|} \text{.}\label{eq:cst4}
	\end{align}
\end{subequations}
Constraint (\ref{eq: singleflowcon}) ensures that up to one flow can enter and  exit node $i$. 
\subsection{Strategic Content Placement Use Case}
In this subsection, we discuss the potential application of the trajectory design to the content caching problem. We assume that each area contains users that request popular file content. 
Furthermore, we assume that the UAV carries a cache in which we can cache popular file content and deliver the files to the users. However, the capacity of the cache is limited and the time flight of the UAV is limited as well. Therefore, we should jointly decide the trajectory and the file placement. Then, in problem \eqref{optproblem: trajdesign}, the reward would be the product of demand times the popularities of the files at each visited location.

\section{Algorithm for UAV Trajectory Design}

%

\subsection{Orienteering}
First, we characterize the complexity of this problem by drawing an analogy from the \emph{Orienteering Problem} (OP) \cite{golden1987orienteering}, a sport in which starting and ending points are specified in a forest along with other locations (checkpoints) with associated scores for visiting. Boyscouts must travel from the starting to the ending point before a certain deadline expires, and on their way they seek to visit a subset of the locations that maximizes the total collected score. Consider, in the OP problem, that $\lambda_{i}$ are the boyscout rewards collecting from each checkpoint $i$, $d_{ij}$ is the travel distance between the checkpoints $i$ and $j$, and $D_{\text{max}}$ is the total time frame available for waypoint collection. Then, there is an 1-1 mapping between the UTD and OP problem. Since the OP problem is NP-hard and particularly APX-hard, we get the following result.\\

\noindent\textbf{Corollary 1.} \textit{The UTD problem in \eqref{optproblem: trajdesign} is NP-hard, and particularly, APX-hard. \footnote{
		It is APX-hard because it has a 2+e poly-time algorithm, but it has no Polynomial-time Approximation Scheme (PTAS), i.e., it cannot be approximated to within any constant larger than 1 \cite{chekuri2012improved}. 
}}\\

\noindent\textbf{Remark. }The authors in \cite{golden1987orienteering} first defined the Orienteering problem, and show that is NP-hard with a reduction from the \emph{Travelling Salesman Problem}. The work in \cite{blum2007approximation} shows that the Orienteering is APX-hard, i.e., any polynomial time algorithm will fail to approximate the optimal within $\frac{1481}{1480}$ (unless P=NP).
Also, it was provided a 4--approximation using dynamic programming to compute min-excess paths, i.e., paths that achieve a targeted prize by introducing a minimum amount of excess cost. \cite{friggstad2017compact} provides a 3--approximation of the rooted Orienteering problem, based on Linear Programming relaxation and rounding.
Improved guarrantees are also given in  \cite{chekuri2004maximum,chekuri2012improved} where a 2--approximation guarantee is provided using $k$-TSP techniques. 

\vspace{1mm}
\subsection{Subtour elimination: lazy constraints approach}

Note that if the number of nodes is of size $n$, then, there are $2^{n}-2$ subsets of $\mathcal{S}$	of $\mathcal{I}$, excluding $\mathcal{S} = \mathcal{I}$ and $\mathcal{S}=\emptyset$. In order to avoid constructing an exponential number of constraints for each scenario resulting in a formulation that is complex even to state, we include the constraints in ($\ref{cons: subtour}$) in a \textit{lazy fashion}. More specifically, we relax all subtour elimination constraints (SECs) \eqref{cons: subtour} and solve the remaining Integer Linear Program (ILP) by using Gurobi solver\footnote{Note that in order to obtain an optimal solution from the solver, we do not restrict its runtime.}. When the solver finds a feasible solution that satisfies the other constraints, we check the number of edges and determine whether the found solution has disconnected subtours or not. If the number of edges of the shortest tour is equal to the number of visited nodes, the found solution has no subtours, hence it satisfies the subtour elimination constraints (even if we did not require them) and the optimization problem is solved. Otherwise, we add the corresponding subtour elimination constraint that is violated and solve the problem again. We repeat until the found solution has no subtours. \\

\noindent\textbf{Theorem 1.} \textit{Lazy constraints approach  is optimal.}
\begin{proof}
 At each iteration, we find a solution of minimum cost of the relaxed problem. We denote the relaxed minimum cost by $c_{\text{min}}$. $c_{\text{min}}$ is a lower bound to the optimal cost of the original problem $c_\text{opt}$. After the last iteration of SEC approach, we find an optimal solution to the relaxed problem that is actually feasible in the original problem. Hence, it must be $c_\text{min}\geq c_\text{opt}$. Therefore we conclude that $c_\text{min}=c_\text{opt}$.
\end{proof}
We note that this approach provides no guarantees that we will not have to eventually add all subtour elimination constraints (and hence it requires exponentially many steps), however, experience shows that it can be quite efficient in some problems \cite{pferschy2017generating}.

\subsection{UAV trajectory algorithm}
\begin{algorithm}[!t]
	\small
	\caption{UAV trajectory design algorithm}\label{alg}
	\textbf{Input}: Graph$=(\mathcal{I},E,\bm{d})$, budget tour $D_{\text{max}}$, start/end node $s$ \\
	\textbf{Output}: trajectory $\mathcal{T}$\\
	 $t_d\leftarrow 0$  //traversed time\\
	 next\_node $\leftarrow \infty$ \\
	 $\mathcal{P}\leftarrow \left\{s,s\right\}$ //we start and end at the node $s$ \\ 
	 next\_segment $\leftarrow \infty$\\
 \Do{\text{next\_node} $\neq \emptyset$}
	{\If{$(\text{next\_node}<I+1)$}
		{$\ell \leftarrow |\mathcal{P}|$\\
	 	 $m\leftarrow$ next\_segment\\
		 $t_d\leftarrow t_d + d_{m,\text{next\_node}} + d_{\text{next\_node},m+1} - d_{m,m+1}$ \\
		 $\mathcal{P}_{m+2:\ell+1}\leftarrow \mathcal{P}_{m+1:\ell}$	\\
		 $\mathcal{P}_{m+1} \leftarrow \text{next\_node}$ 
  		 } 
	\For{$\forall j \in |\mathcal{T}|-1$}{
		$\ell_c\leftarrow \emptyset$, $\ell_e\leftarrow \emptyset$\\
		\For{$\forall i \in \mathcal{I}$}
			{\If{($i\notin \mathcal{T}$)}
				{$i_{1}\leftarrow \mathcal{T}_{j}$\\
				 $i_{2}\leftarrow i$\\
			 	 $i_{3} \leftarrow \mathcal{T}_{j+1}$\\
		 	 	 \If{$(t_d+d_{j,i}+d_{i,j+1}-d_{j,j+1}\leq D_{\text{max}})$}
		 	 	 	{$\ell_c \leftarrow \ell_c \cup i $ //local candidates \\ 
		 	 	 	 $\ell_e \leftarrow \ell_e \cup \left\{t_d+d_{j,i}+d_{i,j+1}-d_{j,j+1}\right\}$ //add extra distance
	 	 	 	 }}
	 	 	 }
 	 	 \eIf{$(\ell_c = \emptyset)$}{
 	 	  $g_{c_{j,:}} \leftarrow \left\{I+2,0\right\}$ // we didn't find any candidate
 	 	 }
 	 	 {$u_{k} \leftarrow \frac{\lambda_{ k}}{\ell_{e_{k}}}\text{, } \forall k \in \ell_c$ // Score \\
 	 	  $\ell_{c_{max}} \leftarrow \text{find the node with  the maximum utility value}$\\
  	  	  $g_c\leftarrow \ell_{c_{max}} \cup max(u) $ //assign the maximum utility value and the corresponding node}
 		}
 	next\_node $\leftarrow \emptyset$, next\_segment $\leftarrow \emptyset$\\
 	\If{$(max(g_{c_{:,2}}>0))$}
 	{ //we found some candidates\\
 	  $u \leftarrow g_{c_{:,2}}$//assign all rows of the second column\\
 	  next\_segment $\leftarrow$ \text{find node with the maximum utility}\\
 	  next\_node $\leftarrow  g_{c_{\text{next\_segment,1}}} $}
}
\textbf{return} $\mathcal{T}$
\end{algorithm}
\begin{table*}[t!]	\caption{The case with $80$ nodes. Execution time. Optimal solution vs greedy algorithm.}
	\centering
	\begin{tabular}{ | c | c | c |  c |  c |  c |  c |  c | c |}
		\hline
		$D_{\text{max}}$ (min) & 2 & 4 & 6 & 8 & 10 & 12 & 14 & 16 \\ 
		\hline
		Solver & 103.22 sec & 377.95 sec & $>2$h & $>2$h & $>2$h  & $>2$h & $>2$h  & 41 sec \\ 
		\hline
		Greedy & 0.35 sec & 0.27 sec & 0.25 sec & 0.25 sec & 0.25 sec & 0.39 sec & 0.31 sec & 0.31 sec\\ 
		\hline
	\end{tabular}
	\label{table: extime}
\end{table*}%
\begin{figure*}[t!]
	\begin{center}
		\begin{tabular}{ccc}
			\begin{subfigure}{0.235\textwidth}\centering\includegraphics[scale=0.25]{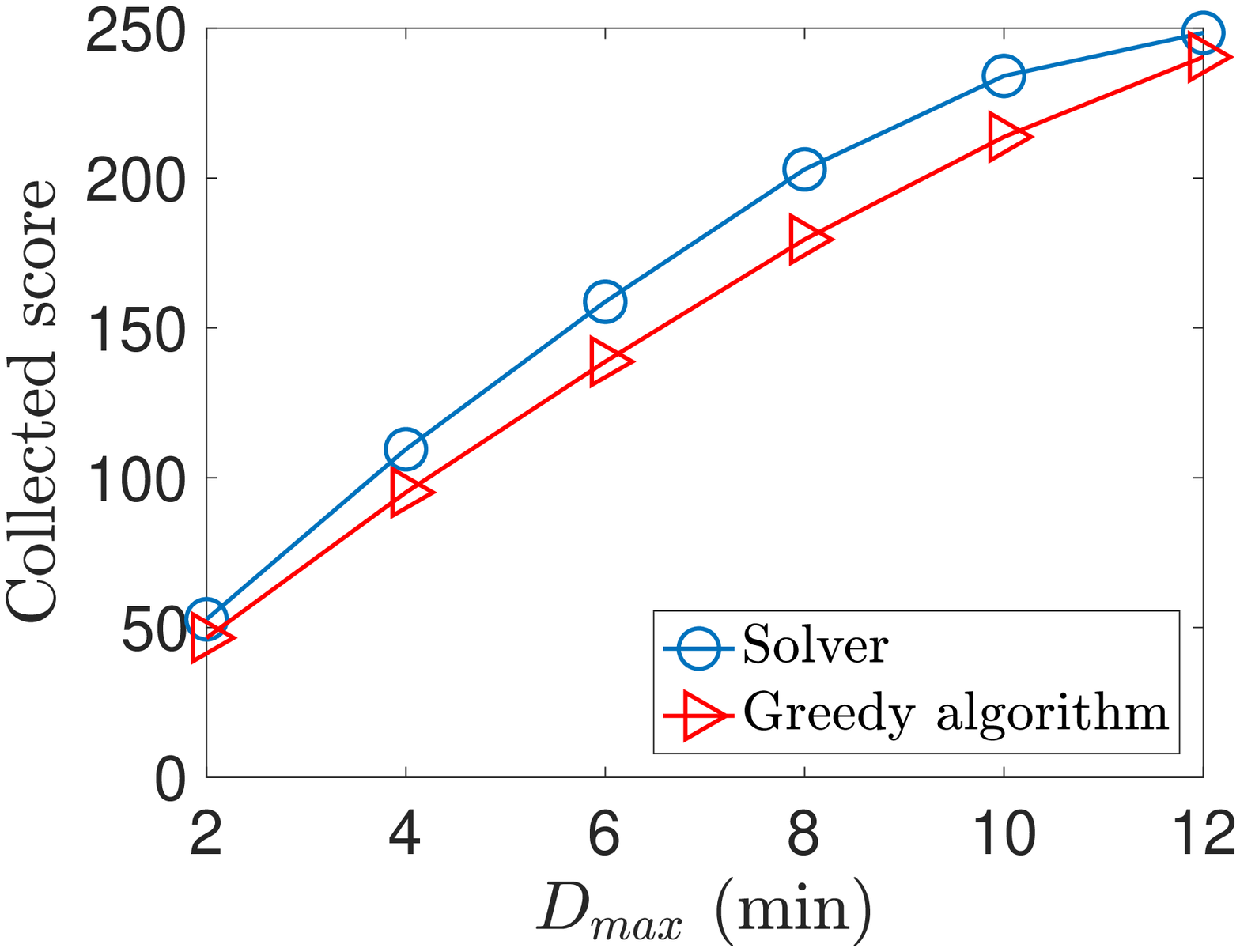}\caption{Collected score by the UAV.}\label{Fig: Score}\end{subfigure}&
			\begin{subfigure}{0.248\textwidth}\centering\includegraphics[scale=0.25]{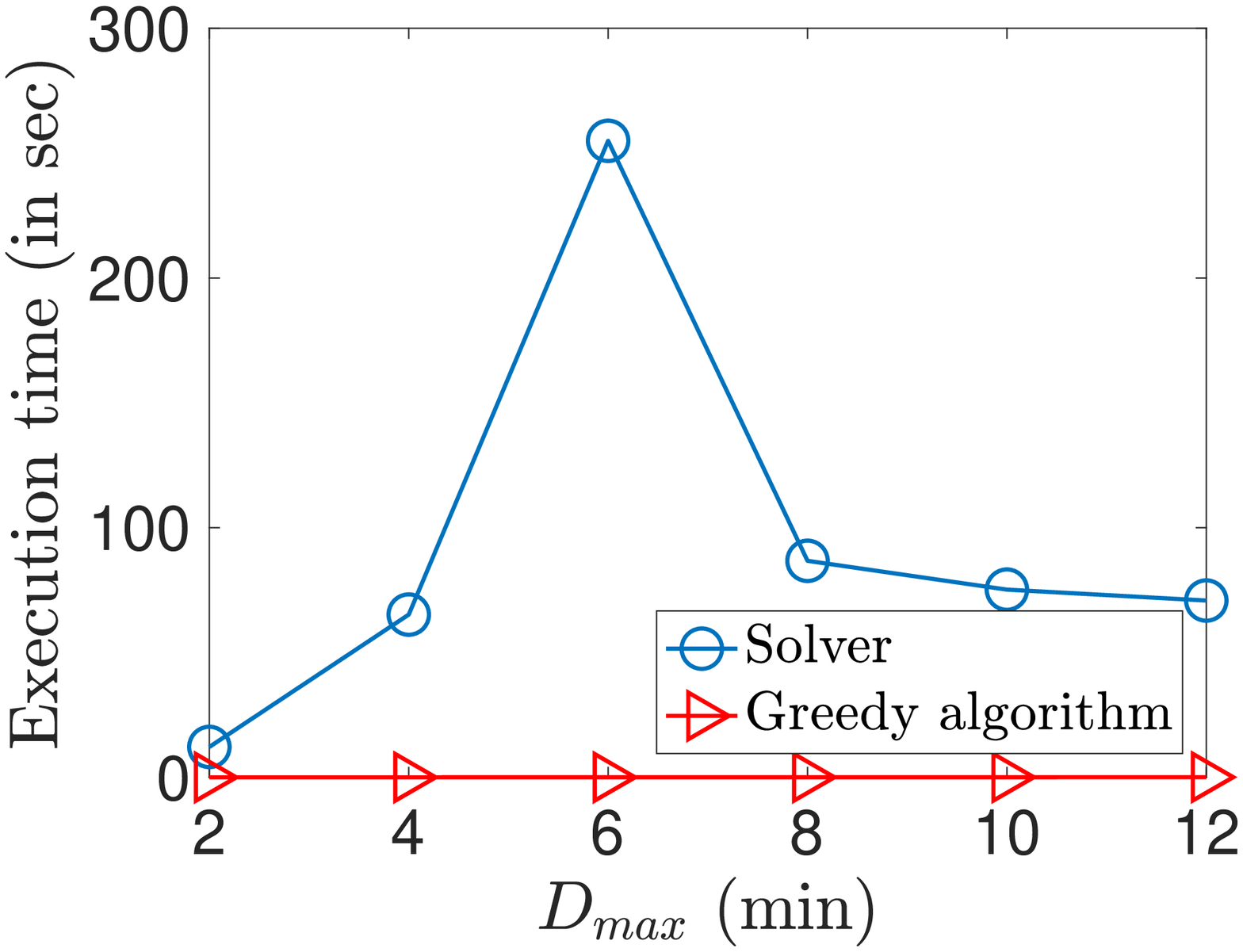}\caption{Execution time.}\label{Fig: ExTime}\end{subfigure}
			\begin{subfigure}{0.248\textwidth}\centering\includegraphics[scale=0.25]{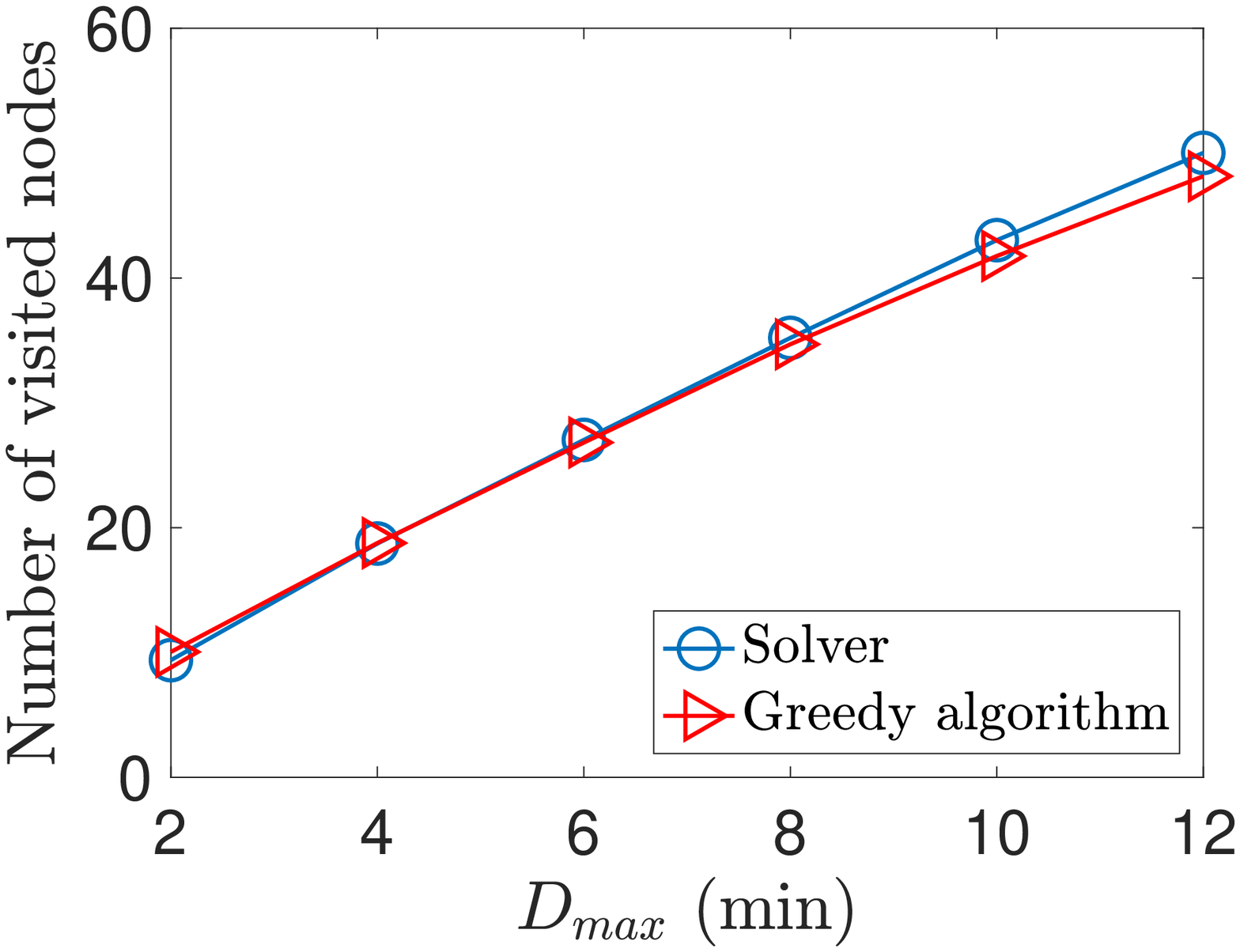}\caption{Number of visited nodes.}\label{Fig: VisitedNodes}\end{subfigure}
			\begin{subfigure}{0.23\textwidth}\centering\includegraphics[scale=0.25]{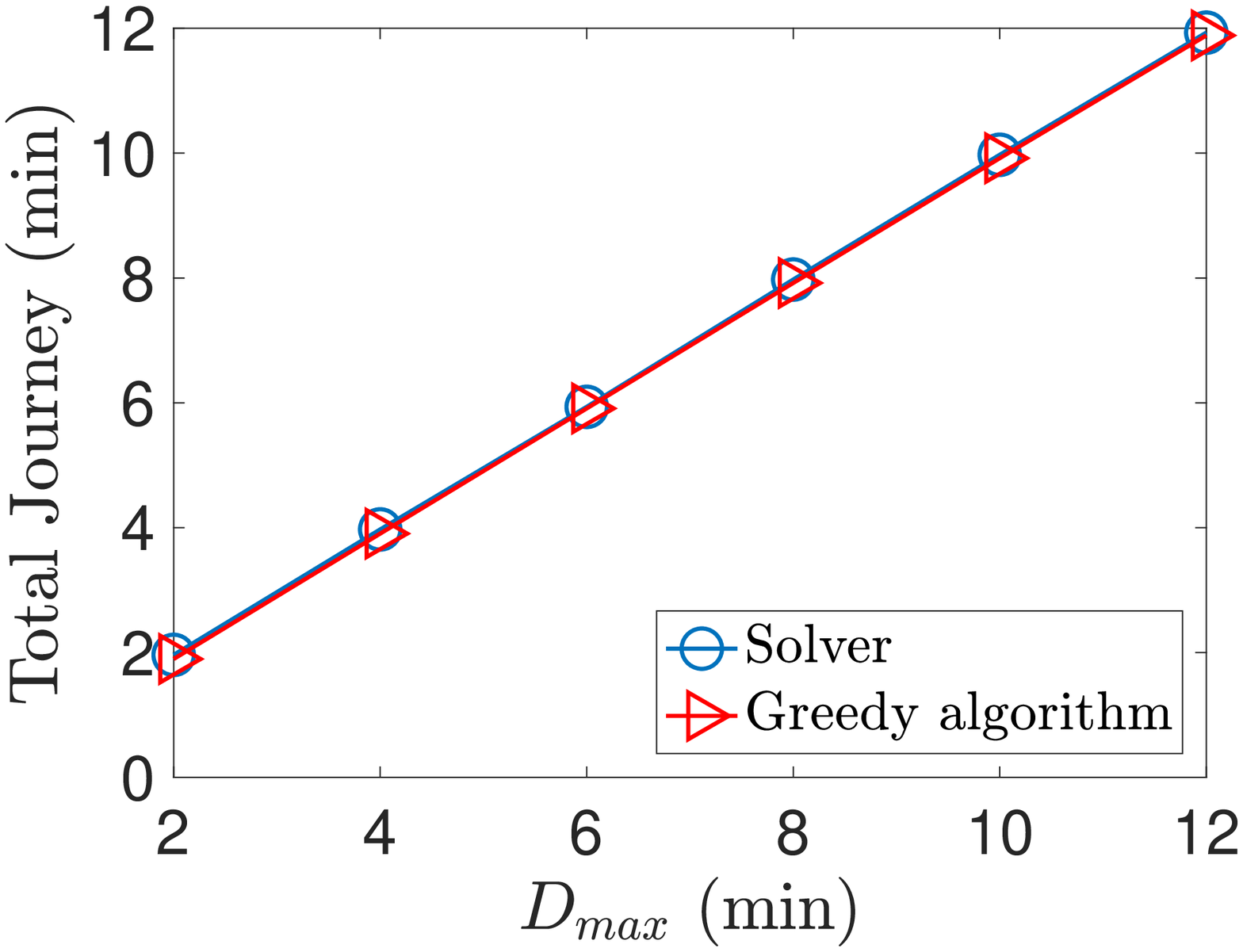}\caption{Total journey time.}\label{Fig: TotalJourney}\end{subfigure}\\[2\tabcolsep]
		\end{tabular}  
	\end{center}
	\caption{The case with $60$ nodes. Optimal solutions vs greedy algorithm.}\label{Fig: results}
\end{figure*}	
Even with the lazy constraints approach, the OP problem is APX-hard, and as the instance of the problem increases, the solver will take too long to return a solution (if ever). In order to have a solution in reasonable time, we propose a heuristic  that is described in Algorithm \ref{alg}. Our approach is inspired from a recent study that proposes touristic itineraries on Google maps \cite{friggstad2018orienteering}. Although our algorithm provides no guarantees, it follows the ideas of the knapsack relaxation, i.e., greedily adding waypoints that maximize the efficiency ratio $\frac{\text{reward}}{\text{added time}}$.

The algorithm builds a trajectory by progressively adding waypoints considering: (i) the feasibility of the tour (step 21), (ii) the cost efficiency of a waypoint addition by means of ratio of reward/added travel time (step 23). Specifically, we begin with the origin and add the waypoint $a$ that maximizes the ratio reward/distance (step 28). At this point the trajectory is simply origin $\rightarrow$ $a$ $\rightarrow$ origin. Next, for every hop in the trajectory, we find the maximal node that if added in the hop, it will maximize the ratio reward/added distance (step 34). Hence it could be $o\rightarrow b \rightarrow a \rightarrow o$ or $o\rightarrow a\rightarrow b \rightarrow o$. Specifically for the second step, there is symmetry and both solutions will be equal. But for the following steps, every hop results in a possibly different maximal waypoint, and we must select the best. At every step of the way, a waypoint can be added only if the new total travel time does not exceed our constraint. When no such node can be found (step 36), our heuristic has converged.

\section{Simulation Results}
We consider that the velocity of the UAV is equal to $70$ km/h\footnote{ \url{https://www.drone-world.com/dji-phantom-4-specs/}.} and  $100$ different topologies with $50$ nodes each. The location of each node is randomly generated according to a normal distribution that takes values in $[-1,1]$. For each topology, we generate score $\lambda_{i}$ for each node. Each score takes values according to a normal distribution in $[0,10]$. We consider that the UAV always starts from $(0,0)$ location points. Optimal and suboptimal trajectories are designed by the solver and algorithm, respectively, for each topology. Then, we take the average of the collected score, execution time, number of visited nodes, and total journey over  the topologies. We repeat for different values of $D_{\text{max}}$. We obtain the optimal solution by using \textit{Gurobi} software.

In Fig. \ref{Fig: Score}, we compare the collected score for the solution provided by the solver and algorithm. We observe that the score collected  by greedy algorithm solution is very close to the optimal one. The algorithm utilizes the available budget in an efficient way, as shown in Fig. \ref{Fig: VisitedNodes} and Fig. \ref{Fig: TotalJourney}. 
The algorithm needs less than $1$ sec to provide an approximate solution, as shown in Fig. \ref{Fig: ExTime}, that is important when we have a large system to solve or need to run the routine multiple times within another algorithm. On the other hand, the solver needs more than $1$ min to provide an optimal solution, and as $D_{\text{max}}$ increases its runtime increases dramatically. However, we observe that the runtime of the solver decreases after a certain point, as shown in Fig. \ref{Fig: ExTime}. The constraint that affects the number of the trajectory options is \eqref{cons: deadline}, i.e., the flight time budget of the UAV. As the time flight budget increases, the number of nodes that can be visited without violating the constraint, approaches the number of the nodes that cannot be visited as shown in Fig. \ref{Fig: VisitedNodes}. Therefore, the trajectory options increase and the solver needs more time to provide the optimal solution. However, when $D_{\text{max}}$ is greater than $6$ min, we observe that the number of visited nodes is greater than the number of not visited ones. For example, for $D_{\text{max}}=8$ min, the UAV visits $35$ nodes out of total $60$ nodes, as shown in Fig. \ref{Fig: VisitedNodes}. Therefore, it is easier now for the solver to find an optimal solution. To give a better intuition on this, consider that the flight time budget is infinite. Then, the solution is trivial; visit all the nodes without taking into account the order. The order does not affect the value of the objective function.

Additional results are provided in Table \ref{table: extime}, for a larger topology with $80$ nodes. We see that for some cases, the solver needs more than $2$ h to provide the solution\footnote{We set up the program to stop after $2$h of waiting period.}. On the other hand, our proposed algorithm can provide an approximate solution in reasonable time for arbitrary number of nodes. 
\section{Conclusions}
In this paper, we study the trajectory design problem of a UAV that flies over multiple areas and collects the corresponding scores. We formulate an optimization problem in order to maximize the collected score over multiple geographical locations. We show that the problem is equivalent to the Orienteering Problem from operation research, and therefore it is APX-hard. We then provide a fast heuristic algorithm, and a simplified MIP approach and compare their performance. Simulation results show that the algorithm performs well and provides solutions for the cases where the solver collapses. The proposed UAV trajectory design problem can be applied for tactical network and strategic content caching applications.
\bibliographystyle{IEEEtran}
\bibliography{mybib_v13}
\end{document}